\documentclass[12pt,a4paper]{article}
\usepackage{graphicx}

\begin{document}

\title{Form Factor Fit for $e^+e^-\rightarrow \pi^+\pi^-\pi^+\pi^-$}
\author{Y. Weng\thanks{e-mail:wengy@ihep.ac.cn}, H. Hu\\
\emph{Institute of High Energy Physics, Beijing} 100049,
\emph{China}}
\date{\today}
\maketitle
\begin{abstract}
The cross section of $e^{+}e^{-}\to \pi^{+}\pi^{-}\pi^{+}\pi^{-}$
has been measured by \emph{BABAR} collaboration. We apply the
theoretical cross section deduced from the extended VMD (Vector
Meson Dominance) model to fit these experimental data. The relevant
parameters and the isovector form factor are obtained.
\end{abstract}

%

\section{Introduction}
~~~~Since 1988, $\rho(1600)$ was replaced by two new resonances:
$\rho(1450)$ and $\rho(1700)$ in PDG. It is suggested that this
possibility can be explained by a theoretical analysis on the
consistency of $2\pi$ and $4\pi$ electromagnetic form factors.
Furthermore, detailed experimental data on the cross section of
$e^{+}e^{-}\to \pi^{+}\pi^{-}\pi^{+}\pi^{-}$ make possible the
accurate determination of the parameters of $\rho$-meson and its
radial recurrencies. Therefore it is meaningful to fit these
parameters simultaneously and compare current theoretical model with
the experimental data.
Recently, VMD model has been developed, which not only includes the
contribution of the lowest-lying vector-mesons, but also includes
those of their high-mass recurrencies [1][2]. The experimental
results of the cross section of $e^{+}e^{-}\to
\pi^{+}\pi^{-}\pi^{+}\pi^{-}$ can be used to demonstrate the above
theoretical model. This paper aims to adopt the cross section of
$e^{+}e^{-}\to \pi^{+}\pi^{-}\pi^{+}\pi^{-}$ derived from the
extended VMD model to fit the experimental data from \emph{BABAR}
collaboration.

\section{Expression of the cross section}
~~~~The cross section of $e^{+}e^{-}\to
\pi^{+}\pi^{-}\pi^{+}\pi^{-}$ can be described generally in terms of
the extended VMD model[1] which takes into account the mixing of the
resonances $\rho(770)$, $\rho^{,}_{1}$, $\rho^{,}_{2}$ in the frame
work of the field theory-inspired approach based on the summation of
the loop corrections to the propagators of the unmixed states. The
cross section can be written as follows \cite{php1}\cite{php2}
\begin{equation}
\sigma(s)=\frac{(4\pi\alpha)^2}{s^{3/2}} |F_{\rho^0\pi^+\pi^-}(s)|^2
 W_{\pi^{+}\pi^{-}\pi^{+}\pi^{-}}(s),
\end{equation}
in which $s$ is the total center-of-mass energy squared,
$\alpha\approx1/137$ is the fine structure factor, and
$W_{\pi^{+}\pi^{-}\pi^{+}\pi^{-}}$ is the final state factor.

According to the vector current conservation, the relation
\begin{math}
g_{\rho^{0}\rho^{0}\pi^{+}\pi^{-}}=2g^2_{\rho\pi\pi}[1]
\end{math}
can be considered as a guide for the corresponding coupling
constant. So the expression of the isovector form factor can be
written as
\begin{eqnarray}
F_{\rho^0\pi^+\pi^-}(s)&=&\left(\begin{array}{c@{\:,\:}c@{\:,\:}c}\frac{m^2_\rho}{f_\rho}&\frac{m^2_{\rho^,_1}}{f_{\rho^,_1}}&\frac{m^2_{\rho^,_2}}{f_{\rho^,_2}}
\end{array}\right)G^{-1}(s)\nonumber \\
&\times&
\left(\begin{array}{c}2g^2_{\rho^0\pi\pi}\\g_{\rho^,_1\rho^0\pi^{+}\pi^{-}}\\g_{\rho^,_2\rho^0\pi^{+}\pi^{+}}\end{array}
\right),
\end{eqnarray}
in which the leptonic coupling constants $f_{\rho_i}$ is

\begin{equation}
\Gamma_{\rho_{i}e^+e^-}=\frac{4\pi\alpha^2}{3f^2_{\rho_i}}m_{\rho_i}.
\end{equation}

The matrix of inverse propagators is
\begin{equation}
G(s)=\left(\begin{array}{ccc}D_{\rho}&-\Pi_{\rho\rho^,_1}&-\Pi_{\rho\rho^,_2}\\-\Pi_{\rho\rho^,_1}&D_{\rho^,_1}&-\Pi_{\rho^,_1\rho^,_2}\\
-\Pi_{\rho\rho^,_2}&-\Pi_{\rho^,_1\rho^,_2}&D_{\rho^,_2}\end{array}\right).
\end{equation}
It consists of the inverse propagators of the unmixed states
$\rho_i=\rho(770)$, $\rho^,_1$ and  $\rho^,_2$,
\begin{equation}
D_{\rho_i}\equiv
D_{\rho_i}(s)=m^2_{\rho_i}-s-i\sqrt{s}\Gamma_{\rho_i}(s),
\end{equation}
where
\begin{eqnarray}
\Gamma_{\rho_i}(s)&=&\frac{g^2_{\rho_i\pi\pi}}{6{\pi}s}q^3_{\pi\pi}\nonumber\\
&+&\frac{g^2_{\rho_i\omega\pi}}{12\pi}(q^3_{\omega\pi}+q^3_{K*K}
+\frac{2}{3}\langle q^3_{\rho\eta}\rangle)\nonumber \\
 &+&\frac{3}{2}g^2_{\rho_i\rho^0\pi^+\pi^-}W_{\pi^+\pi^-\pi^+\pi^-}(s)\nonumber\\
 &+&g^2_{\rho_i\rho^+\rho^-}W_{\pi^+\pi^-\pi^0\pi^0}(s),
\end{eqnarray}
and the nondiagonal polarization operators $\Pi_{\rho_i\rho_j}=\rm
Re\Pi_{\rho_i\rho_j}+i\rm Im\Pi_{\rho_i\rho_j}$. The real parts are
still unknown and may be supposed to be some constants, at the same
time, the imaginary parts could be deduced from the unitarity
relatoin as
\begin{eqnarray}
\rm
&&Im\Pi_{\rho_i\rho_j}=\sqrt{s}[\frac{g_{\rho_i\pi\pi}g_{\rho_j\pi\pi}}{6{\pi}s}q^3_{\pi\pi}\nonumber\\
&+&\frac{g_{\rho_i\omega\pi}}{12\pi} (q^3_{\omega\pi}+q^3_{K^*K}
+\frac{2}{3}\langle q^3_{\rho\eta}\rangle)\nonumber \\
&+&\frac{3}{2}g_{\rho_i\rho^0\pi^+\pi^-}g_{\rho_j\rho^0\pi^+\pi^-}W_{\pi^+\pi^-\pi^+\pi^-}(s)\nonumber
\\ &+&
g_{\rho_i\rho^+\rho^-}g_{\rho_j\rho^+\rho^-}W_{\pi^+\pi^-\pi^0\pi^0}(s)].
\end{eqnarray}

\section{The procedure of form factor fit}
~~~~The cross section of $e^+e^-\to \pi^+\pi^-\pi^+\pi^-$ has been
studied and measured by many experiments. \emph{BABAR} collaboration
has measured this cross section in the center-of-mass energies from
0.6 to 4.5 GeV, considering a hard photo radiated from the initial
state. Because of the wide energy range and the relatively small
errors, we use the expression described in Eq .(1) to fit the
undressed (without vacuum polarization) cross sections \cite{php3}.

The fit is carried out through minimum $\chi$-square
\begin{equation}
\chi^2(\alpha)=\sum^n_{i=1}(\frac{\sigma_i-\sigma_{the}(x_i,\alpha)}{\Delta\sigma_i})^2,
\label{chi2}
\end{equation}
where $\sigma_i$ is the experimental cross section;
$\sigma_{the}(x_i,\alpha)$ is the corresponding theoretical
expressions; $\alpha$ is the vector of free parameters which are
required to be fitted; and $\Delta\sigma_i$ is the measurement error
of the individual cross section.

Then Minuit package \cite{php4} has been chosen as the tool to find
the minimum value of the multi-parameter function Eq .(8) and
analyzes the shape of this function around the minimum. The Minuit
processor MIGRAD which is considered to be the best minimizer for
nearly all functions is used to perform the fit. The fit results are
successful for describing the shape of the cross section and the
form factor of $e^{+}e^{-}\to \pi^{+}\pi^{-}\pi^{+}\pi^{-}$. A good
minimum has already been found and the parameter errors, taking into
account the parameter correlations, have been calculated. But it
seems difficult to use MINOS, another Minuit processor, to calculate
the the errors taking into account both parameter correlations and
non-linearities. As we know, Eq .(1) consists of an integral through
three dimensions (taking into account initial state radiative
correction) and fit for $\chi^2$ is sensitive to some parameters. As
a result, the possibility of calculating the parameter errors by
MINOS is reduced by these effects.

The fit parameters are $m_{\rho_i}, g_{\rho_i\pi^+\pi^-},
g_{\rho_i\rho^0\pi^+\pi^-},$ $g_{\rho_i\rho^+\rho^-}, f_{\rho_i},
\rm Re\Pi_{\rho_i\rho_j}.$ The mass of $\rho_i$ have been measured
relatively well, so in the fit they are fixed to the values in the
PDG \cite{php6}: $m_{\rho_0}=0.7758$GeV, $m_{\rho^{,}_1}=1.465$GeV,
$m_{\rho^{,}_2}=1.720$GeV. Meanwhile, according to the analysis in
\cite{php1}, the real parts of nondiagonal polarization operators
$\rm Re\Pi_{\rho\rho^,_1}$ and $\rm Re\Pi_{\rho\rho^,_2}$ are set to
zero. The parameter $g_{\rho^0\omega\pi}$ has been obtained from the
measurement of $e^+e^-\to \omega\pi$ and its value is fixed to 14.3
\cite{php8}. The physical strong coupling constant
$g_{\rho^0\rho^+\rho^-}$ is set to 6.05 according to \cite{php7}.
$f_{\rho^0}$ is calculated from Eq .(3) and it is fixed to 5.1.
Considering $g_{\rho^{0}\rho^{0}\pi^{+}\pi^{-}}=2g^2_{\rho\pi\pi}$
and $g_{\rho^0\rho^+\rho^-}=g_{\rho^0\pi^+\pi^-}$ \cite{php1}, we
set $g_{\rho^{0}\rho^{0}\pi^{+}\pi^{-}}$ and
$g_{\rho^0\rho^+\rho^-}$ to be 73.205 and 6.05 respectively.

\section{The fit results and the discussion}
~~~~The values of the parameters and their errors in the fit are
presented in Table 1. The cross section and the form factor of
$e^{+}e^{-}\to \pi^{+}\pi^{-}\pi^{+}\pi^{-}$ measured by
\emph{BABAR} and the curves predicted by the extended VMD model are
shown in Figure 1 and 2.

Compared with the preceding work by N.N.ACHASOV who used diverse
experimental data from different channels to carry out the fit
respectively, our work focuses on the possibility to fit the
parameters simultaneously. We have tried many groups of initial
input values and obtained lots of agreeable results. The final
values in the Table 1 are the most proper one in our work because
most values of the parameters are in the limit range in \cite{php1}.
Meanwhile, the $\chi^2/n_{d.o.f}$ is near 1. However, there are some
differences between the values from our work and those from the
previous work.

  In Figure2, the dash line only takes into account the
contribution of $\rho(770)$. The VMD model estimate in this case is
[2]
\begin{equation}
\rm
F_{\rho_0\pi^+\pi^-}(s)=\frac{2g_{\rho\pi\pi}m_{\rho}^2}{m_{\rho}^2-s}.
\end{equation}
The fitted lines agree with the experimental data relatively well
above 0.98Gev.

\begin{figure}
\begin{center}
\includegraphics[width=6.0cm,height=6.0cm]{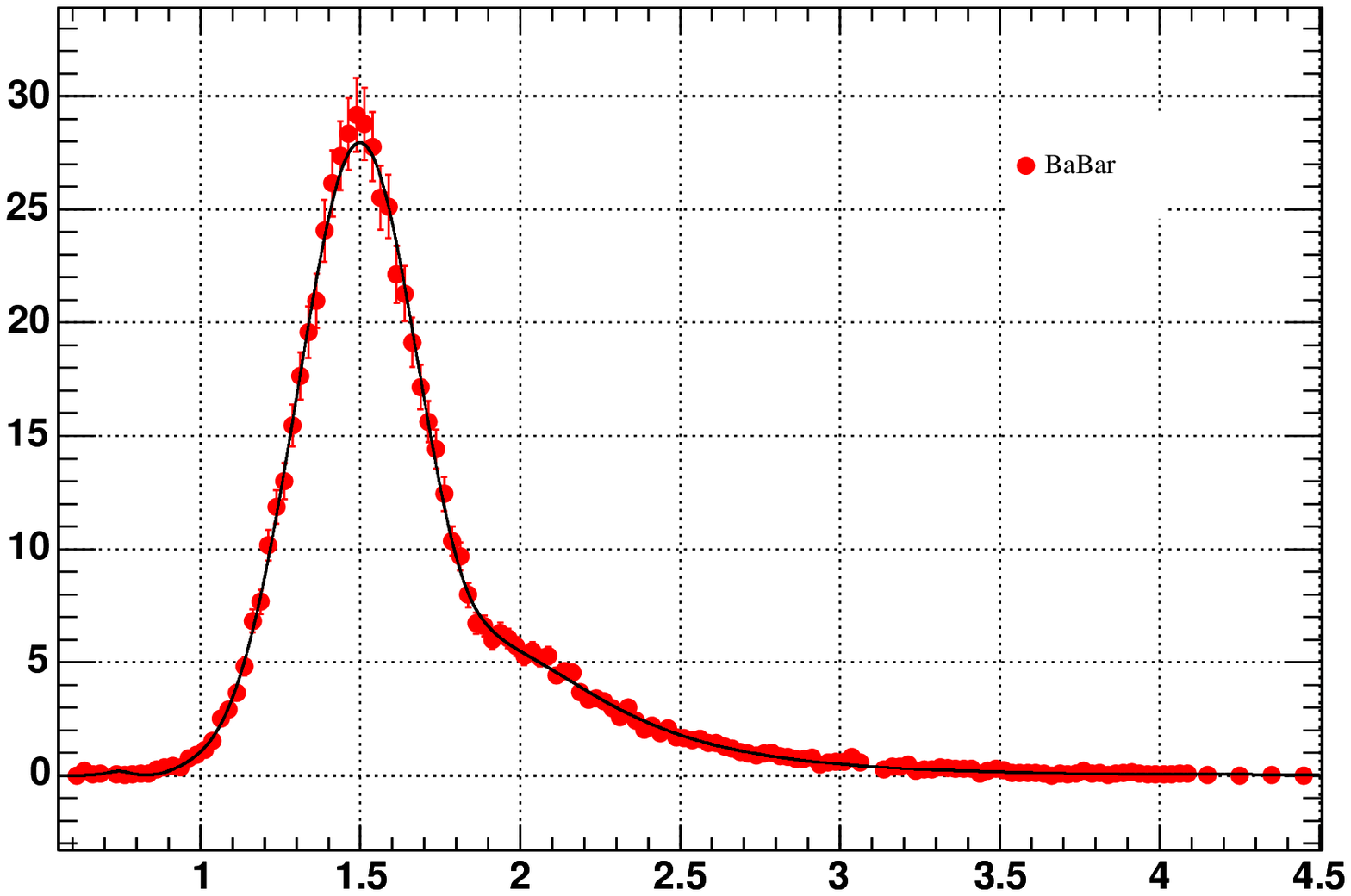}
\put(-70,-5){Ecm (GeV)} \put(-180,50)
{\rotatebox{90}{$\sigma_{e^+e^-\to \pi^+\pi^-\pi^+\pi^-}$(nb)}}
\label{figdavier} \caption{ The energy dependence of the $e^+e^-\to
\pi^+\pi^-\pi^+\pi^-$ cross section obtained from \emph{BABAR} data
in comparison with that from the result of fitting. Total errors are
shown.}
\end{center}
\end{figure}

\begin{figure}
\begin{center}
\includegraphics[width=6.0cm,height=6.0cm]{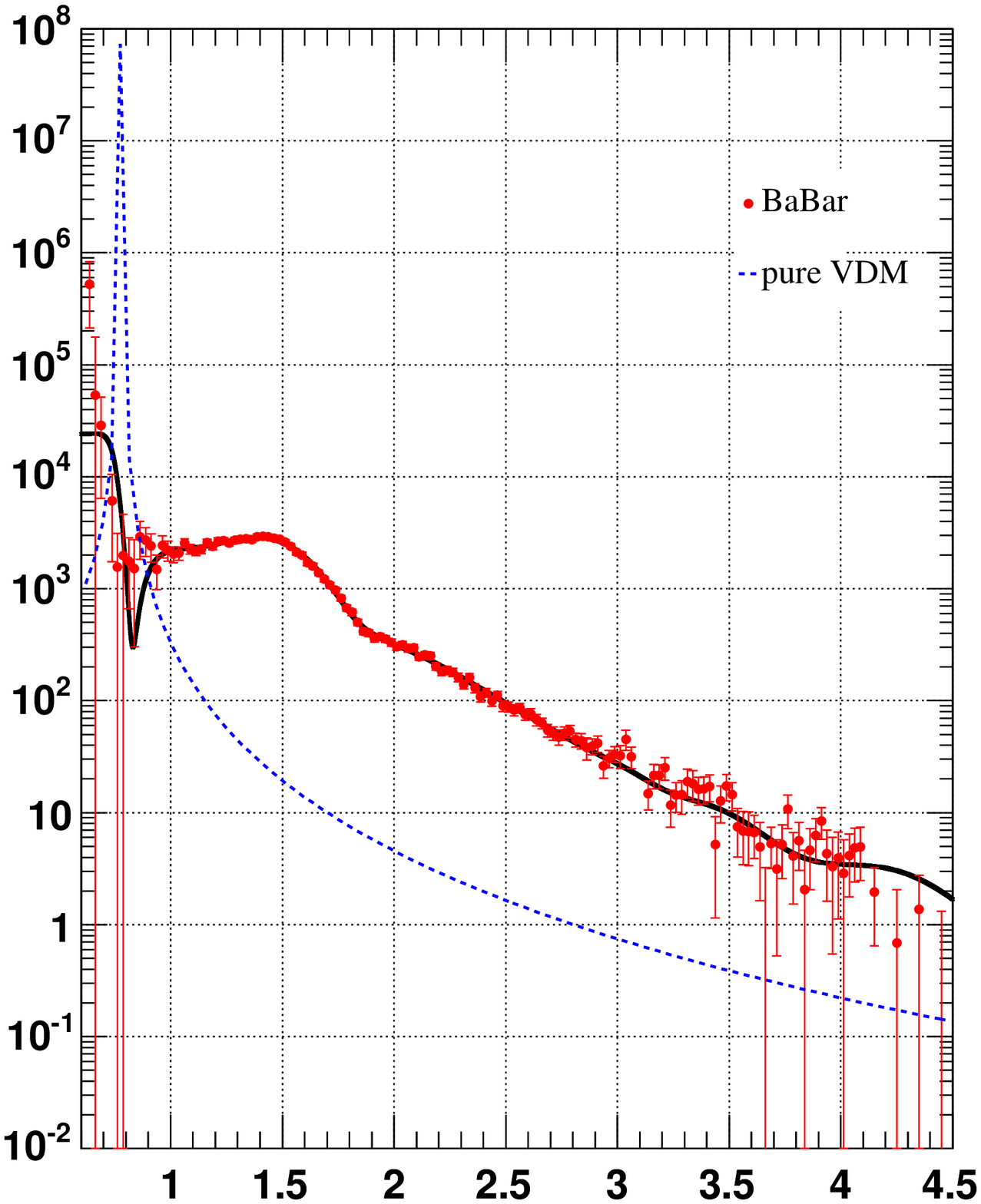}
\put(-70,-5){Ecm (GeV)} \put(-180,106) {\rotatebox{90}{$\left|\rm
F_{\rho^0\pi^+\pi^-}\right|^2$}} \label{figdavier} \caption{The
$\rho\pi^+\pi^-$ form factor squared. The data are recalculated from
the cross section data of \emph{BABAR}. }
\end{center}
\end{figure}

\begin{table}[h]
\caption{The values of the parameters are resulted from fitting with
data \emph{BABAR}.}
\begin{center}
\begin{tabular}{lcc}
\hline\hline Parameter & \emph{BABAR}\\
\hline
$m_{\rho^0}$(GeV) & 0.7758 \\
$m^,_{\rho_1}$(GeV) & 1.4650\\
$m^,_{\rho_2}$(GeV) & 1.7200\\
$f_{\rho^0}$ & 5.1\\
$f_{\rho^,_1}$ &0.442$\pm$0.053\\
$f_{\rho^,_2}$ &0.333$\pm$0.003\\
$g_{\rho^0\pi^+\pi^-}$ & 6.05\\
$g_{\rho^0\omega\pi}$  & 14.3\\
$g_{\rho^0\rho^0\pi^+\pi^-}$ & 73.205\\
$g_{\rho^0\rho^+\rho^-}$ & 6.05\\
$g_{\rho^,_1\pi^+\pi^-}$ &-3.754$\pm$1.078 \\
$g_{\rho^,_1\omega\pi}$  &-4.528$\pm$2.257 \\
$g_{\rho^,_1\rho^0\pi^+\pi^-}$ &229.69$\pm$4.85\\
$g_{\rho^,_1\rho^+\rho^-}$ &-14.454$\pm$0.803\\
$g_{\rho^,_2\pi^+\pi^-}$ &19.241$\pm$1.177\\
$g_{\rho^,_2\omega\pi}$  &29.998$\pm$1.565\\
$g_{\rho^,_2\rho^0\pi^+\pi^-}$ &-410.53$\pm$5.78\\
$g_{\rho^,_2\rho^+\rho^-}$  &-0.520$\pm$1.671\\
$\rm Re\Pi_{\rho\rho^,_1}$  & 0\\
$\rm Re\Pi_{\rho\rho^,_2}$  & 0\\
$\rm Re\Pi_{\rho^,_1\rho^,_2}$ & -6.070$\pm$0.385\\
$\chi^2/n_{d.o.f}$ & 131.89/130 \\ \hline\hline
\end{tabular}
\label{paramt}
\end{center}
\end{table}





\section{Conclusion}

~~~~From the fit, values and errors of the parameters are obtained.
Furthermore, the isovector form factor is also similar to the
previous work. Considering that some parameters may have a strong
correlation and fit for $\chi^2$ is sensitive to some parameters ,
the results of our work may differ from those which are gained by
respectively fitting different experimental data from different
channels.


\end{document}